\documentclass[amsmath,amssymb,amsbsy,prl,twocolumn,superscriptaddress,showpacs]{revtex4}
\usepackage{amsfonts}
\usepackage{amsmath}
\usepackage{amssymb}
\usepackage{bm}
\usepackage[usenames, dvipsnames]{color}
\usepackage{dcolumn}
\usepackage{epsfig}
\usepackage{eucal}
\usepackage{graphicx}
\usepackage[breaklinks,colorlinks = true,linkcolor = red,urlcolor=cyan,citecolor=red]{hyperref}
\usepackage{mathrsfs}
\usepackage{subfigure}
\usepackage{textcomp}
\usepackage{times}
\usepackage{xy}
\usepackage{color}

\usepackage[toc,page]{appendix}

\begin{document}

\title{Exploring the possible origin of spin reorientation transition in NdCrO$_3$}
\author{Hena Das}
\email{das.h.aa@m.titech.ac.jp}
\affiliation{Laboratory for Materials and Structures, Tokyo Institute of Technology, 4259 Nagatsuta, Midori-ku, Yokohama 226-8503, Japan}
\affiliation{Tokyo Tech World Research Hub Initiative, Institute of Innovative Research, Tokyo Institute of Technology, 4259 Nagatsuta, Midori-ku, Yokohama 226-8503, Japan}
\author{Alejandro F. R\'{e}bola}
\affiliation{Instituto de F\'{i}sica Rosario - CONICET, Bv. 27 de Febrero 210 bis,S2000EKF Rosario, Santa Fe, Argentina}
\author{Tanusri Saha-Dasgupta}
\affiliation{Department of Condensed Matter Physics and Material Sciences, S. N. Bose National Centre for Basic Sciences, JD Block, Sector III, Salt Lake, Kolkata, West Bengal 700106, India}

%,S.N. Bose National Centre for Basic Sciences,
%Kolkata 700098, India}
\date{\today}

\begin{abstract}
Spin reorientation transitions and other related magnetic phenomena, which owe their origin to the complex interplay between multiple magnetic sublattices, have long attracted scientific attention both from the perspective of fundamental curiosity and technological applications. In this study, combining first principles calculations together with finite temperature Monte Carlo simulations, we explore the possible origins of reorientation transition of Cr spins in NdCrO$_3$. We construct a NdCrO$_3$ specific magnetic model, consisting of symmetric superexchange interactions between magnetic ions, as well as their magnetic anisotropy. We show that the observed spin reorientation in NdCrO$_3$, arises out of a delicate balance between Nd$-$Cr magnetic exchange interactions, single ion anisotropy of Nd spins, and single ion anisotropy of Cr spins. Moreover, though our model does not take into consideration the effect of anti-symmetric and anisotropic-symmetric magnetic exchanges, the qualitative as well as quantitative agreement of the theoretically derived and the experimentally observed spin-reorientation transition in NdCrO$_3$, confirms the merit of our proposed microscopic model. Our results also propose a hitherto unobserved collective magnetic ordering  in Nd sublattice, which is challenging to detect as it is an extreme low temperature phenomena, therefore calls for further investigations.

\end{abstract}

\maketitle
%%%%%%%%%%%%%%%%%%%%%%%%%%%%%%%%%%%%
%%%%%%%%%%%%%%%%%%%%%%%%%%%%%%%%%%%%%%%%%%%%%%%%%%%%%%%%%%%%%%%%%%%%%%%%%%%%%%%%%%%%%%%%%%%%%%%%%%
\noindent
\section{Introduction}

When multiple magnetic sublattices are formed in a perovskite structure, strong mutual interactions between these superlattices lead to unique magnetic and related phenomena~\cite{RFO-MAG, ref2, ref1, ref5, tokura, Rao, Tokunaga, Zhao, DAS}. The formation of these magnetic sublattices can be attributed to many factors, such as, certain chemical compositions ~\cite{RFO1, RCO1} or  the formation of unique charge-ordered states~\cite{DAS}. In this regard, systems  belonging to the rare-earth (R) transition metal (M) perovskite family, particularly RFeO$_3$ orthoferrites and RCrO$_3$ orthochromites, exhibit a rich variety of magnetic properties resulting from the interplay of two different magnetic sublattices. These systems crystallize in the orthorhombic $Pbnm$ structure. Though the magnetic properties of RCrO$_3$ compounds are similar to those of the isomorphic RFeO$_3$ ones, the former group exhibits a wider range of magnetic phase transitions depending on the characteristic features of the associated rare-earth ions \cite{yamaguchi,MF-RMO}. The transition metal ordering temperatures are smaller by  a factor ranging from two to six  in the orthochromites compared to orthoferrites. For example, the M ordering temperature of NdFeO$_3$ is 690 K~\cite{NFO-TN} as compared to that of NdCrO$_3$, which is reported to be 220 K~\cite{tn1}. This implies weaker M-M interaction in chromites compared to ferrites. Therefore, the M-M interaction in chromites is expected to strongly compete with the other magnetic interactions, such as R-M and R-R interactions. Subtle changes in the relative strength of these magnetic interactions can therefore influence the nature of magnetic behavior. Here, we are interested in NdCrO$_3$, a system which was reported to show multiple phase transitions. However, the microscopic origin of these phenomena is yet to be deciphered. Moreover, for this system the magnitude of the rare-earth transition metal coupling is believed to be at least twice as large compared to its orthoferrite counterpart~\cite{tn2}. This phenomenon is also conjectured to influence the observed multiple phase transitions. However, the underlying microscopic mechanism is still unknown.

%%The rare-earth (R) transition metal (M) perovskite family offers a rich variety of magnetic properties, depending on the choice of R and M atoms, interestingly within the same structural framework of orthorhombic space group.\cite{ref1} In particular, RMO$_3$ compounds are considered as model systems to investigate the interplay of magnetism of two sub-systems, M and R.\cite{ref2} In this context, the most studied systems are orthoferrites and orthochromites.\cite{yamaguchi} Though the magnetic properties of RCrO$_3$ family are similar to those of the isomorphic RFeO$_3$ compounds, there are several important aspects that make the orthochromite better candidates for studying the interplay, (a) the transition metal ordering temperatures are a factor of two to six smaller in the orthochromites compared to orthoferrites. For example, the M ordering temperature of NdFeO$_3$ is 690 K as compared to NdCrO$_3$ which is reported to be 220 K.\cite{tn1} This would imply weaker M-M interaction in chromites compared to ferrites. (b) On the other hand, the magnitude of the rare-earth transition metal coupling is believed to be at least twice larger in orthochromites compared to orthoferrites.\cite{prb} Weakening of M-M interaction and strengthening of M-R interaction is expected to drive the competition between the two fierce in orthochromites, thereby
%making the interplay of magnetism of M and R atoms an important issue.

At high temperature, RMO$_3$ compounds are predominantly G-type antiferromagnets with small canting which results in weak ferromagnetism~\cite{DM1,DM2}. A prominent phenomena in RMO$_3$ compounds is the spin reorientation (SR)~\cite{ref5, MF-RMO}, in which as the temperature is lowered, the direction of the easy axis of the M sublattice magnetization changes from one crystal axis to another. Depending on the direction of the magnetization axis before before and after the SR, six different groups can be identified:\cite{yamaguchi} (I) $G_x$ $\rightarrow$ $G_z$. (II)  $G_x$ $\rightarrow$ $G_y$, (III) $G_x$, no SR (IV) $G_y$, no SR, (V)  $G_z$ $\rightarrow$
$G_y$  (VI) $G_z$ with a nonmagnetic R atom, no SR, where $x$, $y$ and $z$ subscripts refer to easy-axis directions pointing to crystallographic $a$, $b$ and $c$ directions, respectively. In case of RCrO$_3$, system corresponds to R= Ce, Sm or Gd belong to category (I); R = Er belongs to category (II); R= Y, La or Eu belong to category (III); R = Tb, Dy, Ho, Yb, Pr or Tm belong to category (IV); R = Nd belongs to category (V); and R = Lu belongs to category (VI). By contrast, the ferrite series show less variety: system corresponds to R = Pr, Nd, Sm, Tb, Ho, Er, Tm or Yb belong to category (I); for R = Ce or Dy belong to category (II); and for R= Y, La, Eu or Lu belong to category (III); in support of the fact that magnetism in chromites is more intricate and diverse compared to ferrites.

The magnetic properties of orthochromites and orthoferrites have been studied  employing a variety of techniques, namely neutron diffraction and inelastic scattering~\cite{shamir,bertaut}, bulk magnetization and susceptibility measurements on powders and single crystals~\cite{horneich-old}, specific-heat studies on powders and single crystals~\cite{tn2, solidstate}, Mossbauer effect~\cite{jalcom}, and optical-absorption spectroscopy~\cite{horneich}. In comparison, the theoretical studies are limited. 

Within these systems, there are three types of magnetic interactions, M$^{3+}$-M$^{3+}$, M$^{3+}$-R$^{3+}$ and R$^{3+}$-R$^{3+}$, which according to their relative strengths set the following hierarchy: M$^{3+}$-M$^{3+}$ $>$ M$^{3+}$-R$^{3+}$ $>$ R$^{3+}$-R$^{3+}$~\cite{yamaguchi}. Each of these interactions generally consist of the isotropic, anti-symmetric and the anisotropic-symmetric superexchange interactions, apart from the single ion anisotropy of the M$^{3+}$ and R$^{3+}$ ions. This inevitably makes the theoretical study of magnetic properties of RMO$_3$ far from trivial. The most exhaustive study in this respect was carried out by Yamaguchi~\cite{yamaguchi} in 1974 which employs first-order perturbation together with mean-field decoupling to study the spin-reorientation phenomenon in orthochromites and orthoferrites. In this work, the SR phenomena was investigated in the parameter space of a model Hamiltonian comprising of isotropic, anti-symmetric and the anisotropic-symmetric M$^{3+}$-M$^{3+}$, M$^{3+}$-R$^{3+}$ superexchange interactions, and the single-ion anisotropy of the M$^{3+}$ ions. The magnetism of R sublattice was neglected, apart from M-R interaction. While this approach could explain the SR phenomena belonging to category (I) ($G_x$ $\rightarrow$ $G_z$) and category (II) ($G_x$ $\rightarrow$ $G_y$) highlighting role of anti-symmetric and anisotropic-symmetric M$^{3+}$-R$^{3+}$ interactions in SR of category (I) and category(II), the SR  $G_z$ $\rightarrow$ $G_y$ (category (V)), as observed in NdCrO$_3$, could not be  explained.

%The possible cause of the failure of this theory for explaining 
The possible cause for the failure of this theory in explaining
SR in category (V), involving Nd-chromites, was speculated to be the single-ion anisotropy of Nd$^{3+}$, which was neglected in the theory by Yamaguchi~\cite{yamaguchi}, thus leaving the SR in NdCrO$_3$ unsolved.
While the anisotropy energy of Gd$^{3+}$ ion is small enough to be neglected, and that of Dy$^{3+}$ is large enough to be treated
as an Ising spin, the anisotropy energy of Nd$^{3+}$ is intermediate between the two. This may have an important influence in deciding SR in Nd chromites 
%together 
when the Nd-Cr interaction is not negligible. To the best of our knowledge, the interplay between the two has not yet been explored in the context of SR.
%the interplay between the two to the best of our knowledge has
%not been explored in context of SR.

Nd-chromite also appears to be a unique case in context of Nd sublattice magnetism. Below the ordering temperature of the M sublattice, the Nd-M interaction has a tendency to polarize the Nd sublattice following the M sublattice ordering, thus acting as
an effective magnetic field. Evidence of this is observed in Nd nickelates, ferrites and chromites~\cite{ref11-prb}. However, at sufficiently low temperature this coupling may compete with the weak R-R interaction and give rise to cooperative long range ordering in the Nd sublattice. While existence of Nd sublattice ordering
has been established in ferrites~\cite{solidstate}, the stronger Nd-M interaction in chromites puts in doubt/questions the existence of a cooperative magnetic ordering for the Nd sublattice of these systems. This behavior is counter intuitive since the M sublattice ordering temperature is lower in chromites compared to ferrites.
%the strengthening of Nd-M interaction
%in chromites, questions the existence of cooperative ordering of Nd sublattice in chromites.
%On the other hand, this is counter-intuitive as M sublattice ordering temperature is lower in chromite
%compared to ferrite.

All the above issues, make of NdCrO$_3$ a challenging and interesting system which still remains to be understood.
In the present study, we investigate the spin reorientation phenomena in NdCrO$_3$ by considering a spin Hamiltonian which consists of isotropic M$^{3+}$-M$^{3+}$, M$^{3+}$-R$^{3+}$ and R$^{3+}$-R$^{3+}$ superexchange together with single-ion anisotropy of both Nd$^{3+}$ and Cr$^{3+}$ sites. We employ state-of-the-art first-principles density functional theory (DFT) calculations to extract the parameters of the spin Hamiltonian relevant for NdCrO$_3$. The parameters extracted from DFT calculations and based on the experimentally measured crystal structure of NdCrO$_3$ encode the structural and chemical details of the system. Subsequently, the finite temperature magnetic properties of the spin Hamiltonian was obtained by Monte-Carlo simulations considering the spin Hamiltonian. Our results 
%The obtained results 
reveal that such a Hamiltonian is able to capture the spin-reorientation in NdCrO$_3$ correctly. The calculated transition temperatures corresponding to the N\'{e}el ordering of Cr spins and SR are in reasonable agreement
with experimental values. Our study pinpoints the interplay of the Nd-Cr interactions and single-ion anisotropy of Nd and Cr sites in driving this exceptional SR phenomena. Since we are primarily interested in the study of SR and in the role of single-ion anisotropy, this study does not take into account the anti-symmetric and the anisotropic-symmetric superexchange interactions, which would give rise to non-collinear magnetism of Cr spins resulting in a small canting, as reported experimentally.

Moreover, our first principles study combined with Monte-Carlo simulations unravel a yet unreported C-type magnetic ordering in the Nd sublattice, throwing further debate on the nature of cooperative ordering of Nd spins in NdCrO$_3$. Further experimental studies, as well as theoretical ones taking into account the influence of  anti-symmetric and anisotropic-symmetric superexchange interactions are needed to settle the issue conclusively.

The present work underlines the effectiveness of first-principles calculations in capturing the complexity
of rare earth transition metal oxides involved in the delicate balance between magnetic interactions and single ion
%involving delicate balance and interplay of magnetic interactions and single ion
anisotropies of two the magnetic sublattices. It further establishes the power of such approach in providing a
microscopic understanding of SR transition in NdCrO$_3$, phenomenon that still remains unsolved by the theory. 
%which till date has remained unsolved theoretically.

\begin{figure}
\includegraphics[width=1.0\linewidth]{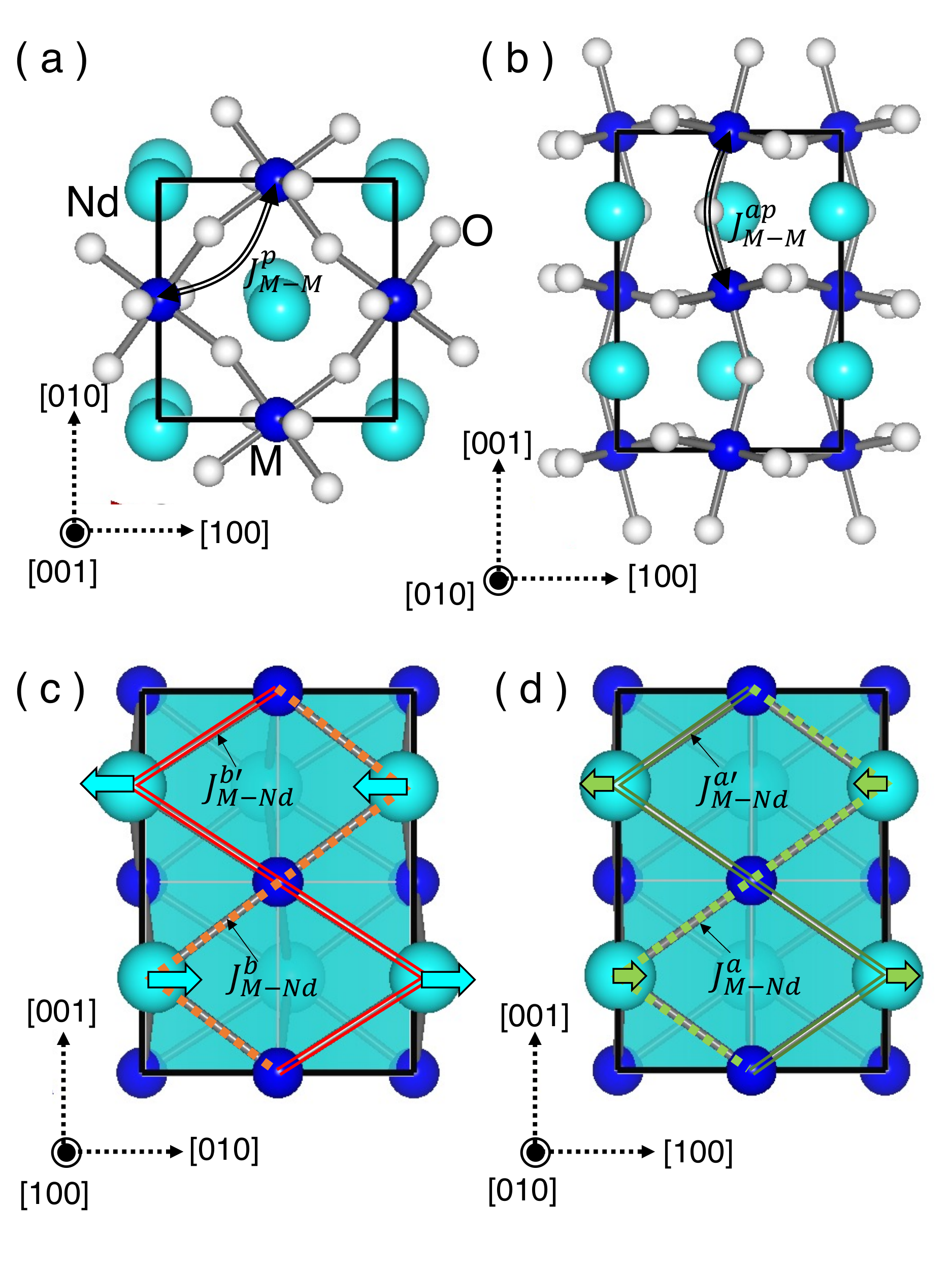}
\caption{Orthorhombic $Pbnm$ crystal structure of NdCrO$_3$. Crystal structure showing the first nearest-neighbor isotropic M-M exchange interactions, in plane  ($J_{M-M}^{p}$) (a) and out of plane ($J_{M-M}^{ap}$) (b). Crystal structure plot exhibiting
 four in-equivalent nn M-Nd exchange interactions, denoted as $J_{M-Nd}^b$ and $J_{M-Nd}^{b\prime}$ (in the crystallographic \textit{bc} plane) (c) \& $J_{M-Nd}^a$ and $J_{M-Nd}^{a\prime}$ (in the crystallographic \textit{ac} plane) (d). The arrows indicate the anti-ferro off-centric displacements of Nd ions.}
\label{NCO-S}
\end{figure}

\section{Computational Details}

Using the experimentally determined $Pbnm$ crystal structures, the symmetric exchange interactions between the magnetic ions, as well as their magnetic anisotropy parameters were estimated through the calculation of total energy of various collinear magnetic configurations employing the density functional theory (DFT) based linearized augmented plane-wave (LAPW) method as implemented in the Wien2k code~\cite{wien2k1,wien2k2}. We considered sixteen magnetic configurations in order to estimate the strength of the symmetric exchange interactions and six magnetic configurations to estimate the magnetic anisotropy parameters of the ions. We used the Perdew-Burke-Ernzerhof (PBE)~\cite{PBE} Generalized Gradient Approximation (GGA) form of exchange correlation functional. The effect of the missing correlation on the transition metal Cr sites (with localized 3$d$ electrons) and rare earth Nd sites (with localized 4$f$ electrons) beyond the GGA was taken into account through supplemented Hubbard $U$ and Hund’s coupling $J_H$ terms using GGA+$U$ method~\cite{DFT+U}. The choice of appropriate $U$ and $J_H$ values, for that matter, is very crucial in quantitative description of the magnetic and electronic structures of strongly correlated insulators like the system in question. Following the values of $U$ estimated by employing the constrained density functional theory (cDFT) method~\cite{cDFT} and previous theoretical studies~\cite{Zhao,U-cr}, we performed our calculations considering $U$ for Cr in the range of 2 - 5.5 eV (0.15 - 0.40 Ryd). On the other hand, we set a 5.5 eV (0.40 Ryd) value of $U$ at the Nd site. This strategy gave us the opportunity to study the effect of the relative strength of $U$ at the Cr and Nd sites on the properties of NdCrO$_3$. Additionally, we  considered a range of values, 0 - 1.0 eV, for $J_H$ parameter, a factor which is found to have a strong influence on the estimated magnetic parameters. We performed our calculations using a plane-wave cutoff of $RK_{MAX}$ = 7 and Monkhorst-Pack $\Gamma$ centred \textit{k}-point mesh of 6$\times$6$\times$4 for $Pbnm$ structure. We also considered 1$\times$2$\times$1 and 2$\times$1$\times$1 supercell structures to estimate R-M magnetic interactions and the corresponding \textit{k}-point meshes were of the dimensions, 6$\times$4$\times$4 and 4$\times$6$\times$4, respectively.

Monte Carlo (MC) simulations were performed on an 8$\times$8$\times$8 cell consisting of 4096 magnetic ions, and considering 10$^9$ MC steps for each temperature using the model Hamiltonian constructed through GGA+$U$ calculations. In Addition, we conducted finite temperature MC simulations further considering a wide variation in magnetic parameter range  of  R-M superexchanges, and single ion anisotropy of both R and M ions beyond the DFT estimated values, to unravel and identify the driving forces behind the complex magnetic behaviour of NdCrO$_3$.
In order to determine the primary collinear G-type AFM phase in the Cr sublattice, we calculated magnetic order parameter defined as
\begin{eqnarray}
[m_a,m_b,m_c]=<{\bf S}_1-{\bf S}_2+{\bf S}_3-{\bf S}_4>/4S
\label{AFM}
\end{eqnarray}
where ${\bf S}_1$ $\rightarrow$ ${\bf S}_4$ represent four Cr ($S$ = 3/2) ions in the $Pbnm$ unit cell. $m_a$, $m_b$ and $m_c$ denote the staggered moments along the crystallographic $a$, $b$ and $c$ axis, respectively. We calculated specific heat as a function of temperature using,
\begin{eqnarray}
C_v(T)=\dfrac{\langle \xi^2\rangle - \langle \xi \rangle^2}{k_B T^2}
\label{Cv}
\end{eqnarray}
where the angles bracket denotes thermal average and $\xi$ represents total energy of the system.

\section{RESULTS}

\subsection{Crystal Structure}

The NdCrO$_3$ compound crystallizes in orthorhombic $Pbnm$ space group, which is the GdFeO$_3$-type distorted perovskite structure with both in-phase and out-of-phase CrO$_6$ octahedral tilting distortion pattern of $a^-a^-c^+$, as shown in Figure~\ref{NCO-S}(a) and (b). The lattice constants of the orthorhombic unit cell were measured as \cite{structure}, $a$ = 5.421 \AA, $b$ = 5.487 \AA and $c$ = 7.694 \AA. The experimentally determined Wyckoff positions corresponding to Nd, Cr, O$_{ap}$ (apical oxygen) and O$_p$ (planar oxygen) ions are $4c$ ($x$, $y$ = -0.009, 0.042), $4b$, $4c$ ($x$, $y$ = 0.089, 0.480) and $8d$ ($x$, $y$, $z$ = -0.287, 0.285, 0.040), respectively. The resultant orthorhombic distortion, measured as $\frac{2 \times (b-a)}{(b+a)}$ is 0.012. In addition to rotation $a^0a^0c^+$ ($M_3^+$) and tilt $a^-a^-c^0$ ($R_4^+$) distortions, Nd ions exhibit anti-ferro off-centric displacements along the crystallographic $a$ and $b$ axes, following the $R_5^+$ (Figure 1(c)) and $X_5^+$ (Figure 1(d)) symmetries, respectively. This leads to the formation of Nd-Cr short bonds (3.165 \AA  ~and 2.293 \AA) and long bonds (3.370 \AA ~and 3.540 \AA), respectively. We considered this crystal structure in order to conduct the first-principles electronic structure calculations and to construct the magnetic model Hamiltonian to perform finite temperature Monte Carlo simulations. Considering the fact that no significant change in orthorhombic distortion was reported below the N\'{e}el temperature ($T_N$)~\cite{structure}, we kept the crystal structure fixed in the temperature range of the Monte Carlo simulations.\\
%
%Our calculations on NdFeO$_3$ were conducted using its Pbnm crystal structure experimentally determined at room temperature[Ref] having lattice parameters $a$ = 5.453 $\AA$, $b$ = 5.581 $\AA$ and $c$ = 7.763 $\AA$. The corresponding Wyckoff positions of Nd, Fe, O$_{ap}$ (apical oxygen) and O$_p$ (planar oxygen) ions are $4c$ ($x$, $y$ = 0.492, 0.548), $4b$, $4c$ ($x$, $y$ = 0.408, 0.023) and $8d$ ($x$, $y$, $z$ = 0.790, 0.776, 0.051), respectively.

\subsection{Model Hamiltonian}

To investigate the magnetic phase transitions in NdCrO$_3$ at finite temperatures we constructed a magnetic model Hamiltonian comprising of isotropic exchange interactions between the magnetic ions and the magnetic anisotropy energies of the magnetic ions. The isotropic exchange component of the Hamiltonian is given by, 
\begin{eqnarray}
H_{SE} = H_{M-M}+H_{M-Nd}+H_{Nd-Nd}
\label{E1}
\end{eqnarray}

Where,
\begin{equation}
\begin{split}
H_{M-M} =  \sum_{ij} J_{M-M}^{ap} \textbf{S}_i\cdot \textbf{S}_j + \sum_{ij} J_{M-M}^{p} \textbf{S}_i\cdot \textbf{S}_j \\+\sum_{ij} J_{M-M}^{nnn} \textbf{S}_i\cdot \textbf{S}_j  
\end{split}
\label{E2}
\end{equation}
represents the nearest neighbor (nn) isotropic exchange interactions between M spins (denoted by \textbf{S}) mediated via apical ($J^{ap}_{M-M}$) and planar ($J^p_{M-M}$) oxygen (cf Figs 1(a) and 1(b)) and their next-nearest-neighbor (nnn) interactions ($J^{nnn}_{M-M}$). 

The most important interactions are between two magnetic sublattices and the associated energy component is given by, 
\begin{equation}
\begin{split}
H_{M-Nd} = \sum_{ij}J_{M-Nd}^b \textbf{S}_i\cdot \textbf{S}^{\prime}_j + \sum_{ij}J_{M-Nd}^{b\prime} \textbf{S}_i\cdot \textbf{S}^{\prime}_j \\+\sum_{ij}J_{M-Nd}^a \textbf{S}_i\cdot \textbf{S}^{\prime}_j +\sum_{ij}J_{M-Nd}^{a\prime}\textbf{S}_i\cdot \textbf{S}^{\prime}_j 
\end{split}
\label{E3}
\end{equation}
which incorporates the effect of the orthorhombic displacement of the Nd ions, resulting in four inequivalent exchange interactions;  $J_{M-Nd}^b $, $J_{M-Nd}^a $, $J_{M-Nd}^{b\prime}$ and $J_{M-Nd}^{a\prime}$ (cf Figs. 1(c) and 1(d)). $\textbf{S}^{\prime}$ denotes Nd spin. The first pair of interactions corresponding to the short Nd-Cr bond connections while the second pair represents the same along the longer Nd-Cr bonds. Here we define a parameter $\gamma$ which denotes the strength of the strongest nn isotropic exchange interaction between two magnetic sublattices $J_{M-Nd}$ relative to the strongest exchange interaction between M spins $J_{M-M}$, \textit{i.e.} $\gamma = \dfrac{J_{M-Nd}}{J_{M-M}}$. 

In addition, we considered the nn isotropic exchange interactions between the Nd spins along the three crystallographic axes, denoted as, $J_{Nd-Nd}^a $,  $J_{Nd-Nd}^b$ and  $J_{Nd-Nd}^c$, respectively. The corresponding energy term is given by,
\begin{equation}
\begin{split}
H_{Nd-Nd} = \sum_{ij} J_{Nd-Nd}^a \textbf{S}^{\prime}_i\cdot \textbf{S}^{\prime}_j+\sum_{ij}J_{Nd-Nd}^b \textbf{S}^{\prime}_i\cdot \textbf{S}^{\prime}_j\\ +\sum_{ij}J_{Nd-Nd}^c \textbf{S}^{\prime}_i\cdot \textbf{S}^{\prime}_j
\end{split}
\label{E4}
\end{equation}

The magnetic anisotropy of the system was modelled by considering the single ion anisotropy (SIA) energies of the M and Nd ions.  The corresponding component of the spin Hamiltonian is given by,
\begin{equation}
\begin{split}
H_{SIA} = \sum_i \left\lbrace E_{M} (S^2_{ix}-S^2_{iy})+D_{M}S^2_{iz}\right\rbrace \\+\sum_i \left[  E_{Nd} \left\lbrace (S^{\prime}_{ix})^2-(S^{\prime}_{iy})^2\right\rbrace +D_{Nd}(S^{\prime}_{iz})^2 \right]  \
\end{split}
\end{equation}
The $x$, $y$ and $z$ directions correspond to the crystallographic $a$, $b$ and $c$ axes, respectively, of the $Pbnm$ structure. For ($E<0,D<0,|E|>|D|$) and ($E<0,D>0$) conditions, the spins tend to orient along the $x$ (crystallographic $a$) axis. On the other hand, ($E<0,D<0,|E|<|D|$) and ($E>0,D<0,|E|<|D|$) denote the preference of spins to orient along $z$ (crystallographic $c$) axis. Finally, ($E>0,D<0,|E|>|D|$) and ($E>0,D>0$) tend to orient the spins along $y$  (crystallographic $b$) axis.  The analysis based on site symmetries of both Cr and Nd ions shows the possible existence of non-zero off-diagonal components of the associated SIA tensors in addition to the non-zero diagonal components. However, in the present study, we have taken only the latter into consideration. 

\begin{figure}
\includegraphics[width=1.0\linewidth]{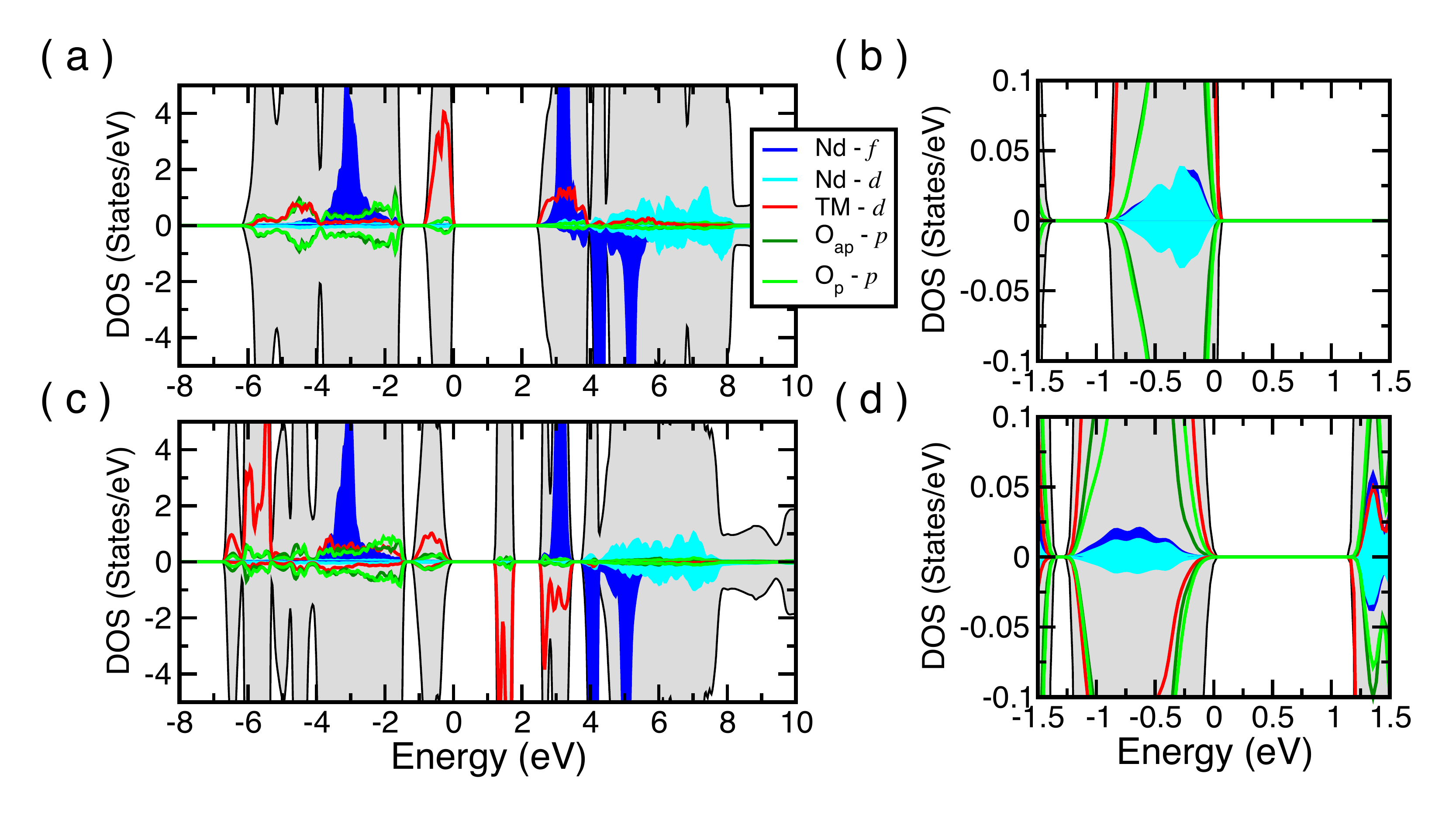}
\caption{(a) and (c) Calculated density of states (DOS) for NdCrO$_3$ and NdFeO$_3$ considering collinear G-type antiferromagnetic order in Cr sublattice as well as Nd sublattice.
  The Nd-Cr and Nd-Fe hybridization are highlighted in zoomed plot (b) and (d). We considered $U$ = 2.2 eV and $J_H$ = 0.3 eV for Cr 3$d$ states and $U$ = 4.5 eV and $J_H$ = 0.3 eV for Fe 3$d$ states. A higher $U$ value of
  5.5 eV and $J_H$ = 0.3 eV for Nd 4$f$ states was used.}
\label{DOS}
\end{figure}

\subsection{DFT Electronic structure}

Figure ~\ref{DOS}(a) shows the electronic structure of NdCrO$_3$ computed within GGA+$U$ formalism and considering a G-type antiferromagnetic order in both the Nd and Cr sublattice. The results presented in the following are for the choice of $U_{Cr}$ = 2.2 eV, $U_{Nd}$ = 5.5 eV and $J_H$ = 0.3 eV, which were found to provide best description of magnetic moments
and exchange interactions vis-à-vis the experimental observations. The system shows insulating behavior with $\sim$ 2.5 eV band gap. The approximate octahedral oxygen environment surrounding Cr ions splits the 3$d$ states into $t_{2g}$ and $e_g$ manifolds. The $t_{2g}$ manifold is filled (empty) in the majority (minority) spin channel for Cr, indicating a nominal 3+ valence state as reported in orthochromites. The electron occupancy of the Nd – 4$f$ orbitals also indicates a 3+ valence state. The values of the calculated spin moments  at Cr and Nd sites are  $\sim$ 2.52 $\mu_B$ and $\sim$ 2.96 $\mu_B$, respectively.
These spin moments provide additional support to the nominal 3+ valences at both transition metal (M) and rare-earth (R) sites. The 4$f$ electrons of Nd are coupled with its 5$d$ electrons by the intra-atomic exchange interaction.  Also,  since the 5$d$ orbitals are spatially extended, they hybridize with the M - 3$d$. This hybridization has been clearly depicted  in the zoomed plot in Figure ~\ref{DOS}(b), which shows a strong overlap between the Nd-5$d$-4$f$ and the Cr - 3$d$ states. For comparative analysis, we also calculated the electronic structure of the G-type ordered magnetic phase of NdFeO$_3$ considering $U$ = 4.5 eV and $J_H$ = 0.3 eV at the Fe 3$d$ states, presented in Fig.~\ref{DOS}(c) and (d). As in the case of orthochromites, in this orthoferrite too, both Fe and Nd ions tend to favor the formation of +3 valence state. Very interestingly, we find that the integrated DOS of Nd-5$d$ in the energy range of -1.5 eV to Fermi level is approximately 2 times higher in NdCrO$_3$ compared to that in NdFeO$_3$, which indiactes that the Nd-Cr hybridization is about 2 times stronger than the Nd-Fe hybridization. This relation is in good line with the experimentally observed relation (based on the specific heat and neutron-diffraction measurements~\cite{tn2} between the values of the mean-field Nd-Cr and Nd-Fe interaction parameters ($n_{M-R}$), where the former is 2.6 times as that of the latter. 

\begin{table}
\caption{Estimated values of isotropic exchange interactions and SIA parameters using $U$=2.2 eV and 5.5 eV for the transition metal Cr 3$d$, and rare-earth 4$f$ states, respectively, and $J_H$ = 0.3 eV.}
\begin{tabular}{c |c c c|c c c c }
\hline
\multicolumn{8}{c}{Isotropic Exchange (meV)}\\
\hline
System & $J_{M-M}^{p}$ & $J_{M-M}^{ap}$ & $J_{M-M}^{nnn}$ & $J_{M-Nd}^b$ &$J_{M-Nd}^{b\prime}$&$J_{M-Nd}^a$&$J_{M-Nd}^{a\prime}$\\
\hline
NdCrO$_3$&4.66&6.01&0.09&2.63&-2.08&0.84&-0.23\\
%NdFeO$_3$&5.63&9.08&-0.52&2.31&-1.94&0.92&-0.40\\
\hline
\end{tabular}
\begin{tabular}{c| c c| c c}
\multicolumn{5}{c}{SIA (meV)}\\
\hline
System & $E_{Nd}$ & $D_{Nd}$& $E_{M}$ & $D_M$\\
\hline
NdCrO$_3$ & -0.40 &-0.10 & 0.003 & -0.01 \\
%NdFeO$_3$ & -0.60& 0.49 & 0.003 & 0.001 \\
\hline
\end{tabular}
\label{T_SE}
\end{table}

\subsection{Estimated Isotropic Exchange Interactions}

We estimated the values of the isotropic exchange interactions between magnetic ions of NdCrO$_3$ using the calculated total energies of sixteen spin configurations and the results are listed in Table ~\ref{T_SE} for $J_H$ = 0.3 eV and the $U$ value of 2.2 eV and 5.5 eV corresponding to the Cr 3$d$ and Nd 4$f$ states, respectively. Additionally, we conducted the calculations considering a range of $U$ and $J_H$ values. The results are summarized in the Supplementary Fig. S1 and S2. We estimated the values of exchange interactions considering spin value of 3/2 for both Cr and Nd. The strongest interaction, $J_{Cr-Cr}$ = 6.01 meV, corresponds to the antiferromagnetic (AFM) nn interaction between Cr spins mediated via the apical oxygens. The Cr spins are also coupled antiferromagnetically in the $ab$ plane, leading to the stabilization of the G-type AFM phase, which is in agreement with the experimental observations~\cite{ref1}. The nnn interactions are comparatively weak and AFM in nature with an average value of $\sim$ 0.09 meV, which is expected to cause weak magnetic frustration in the Cr sublattice. The values of the nn exchange interactions increase linearly with the increase of $J_H$ (see Supplementary Fig. S1(a)).

The interactions between two magnetic sublattices significantly vary with the value of $J_H$, as shown in the Supplementary Fig. S1(b). The orthorhombic distortions result in AFM and FM interactions between the Nd and Cr spins along the shorter and longer bond directions, respectively. We observe that $|J_{Cr-Nd}^b - J_{Cr-Nd}^{b\prime}| > |J_{Cr-Nd}^a-J_{Cr-Nd}^{a\prime}|$, implying that the difference in magnetic interactions due to anti-ferro off-centric displacements of Nd ions along crystallographic $b$ axis is higher than that along the crystallographic $a$ axis. Interestingly, the displacement of the Nd ions along $b$ axis ($Q_b$) is higher in order of magnitude compared to that along the $a$ axis ($Q_a$). Thus, the orthorhombic separation between magnetic interactions directly vary with the amplitude of the off-center displacement of the Nd ions. The strongest interaction corresponds to AFM $J_{Cr-Nd}$ = $J_{Cr-Nd}^b$ = 2.63 meV which gives rise to the relative strength of $\gamma$ = 0.44.
This relative strength varies from 0.25 - 0.89 as we vary the $J_H$ value from 0 to 1.0 eV (see Supplementary Fig. S2). This indicates a strong correlation between the exchange interactions between the two magnetic sublattice and $J_H$. The average exchange interaction ($J_{Cr-Nd}^{avg}$) between two magnetic sublattices is AFM in nature, which is in line with the experimental observation~\cite{shamir}. It is important to point out here, that a similar
exercise carried out for NdFeO$_3$ with $U$ = 4.5 eV and J$_H$ = 0.3 eV at the Fe site, gave a $\gamma$ value of 0.25, a factor of 1.8 smaller compared to NdCrO$_3$, in good agreement with our conclusions on relative Nd-M hybridization between chromites and ferrites from density of states
as well as that concluded from specific heat measurement~\cite{tn2}.

In addition, we also estimated the magnetic interactions between Nd spins, which were found to be weak and AFM in nature along all three crystallographic axes with an average value of 0.02 meV, indicating a G-type magnetic order in the Nd sublattice. However, these interactions are of order of magnitude weaker than the magnetic interactions between two sublattices. %Therefore, in the formation of magnetic order in the Nd sublattice, the latter is expected to dominate.

\subsection{Estimated Single Ion Anisotropy (SIA)}

Next, we calculated the magnetic anisotropy energy of the Nd ions, and the results
 are given in Table ~\ref{T_SE}. In order to decouple the contribution of the Cr spin
 sublattice, we only employ Spin-Orbit (LS) coupling at the Nd spins. As the electronic configuration of R-4$f$ electrons follows Hund's rule, the orbital magnetic moment appears in the presence of spin-orbit coupling (SOC). Once the direction of the 4$f$ orbital moment gets fixed, the direction of the spin moment also gets fixed by the LS coupling. The anisotropy energies were calculated by considering spin configurations oriented along the crystallographic $a$ ($x$), $b$ ($y$) and $c$ ($z$) axes. We observed that, irrespective of the value of $J_H$, the anisotropy energy associated with the orientation of Nd spins along the crystallographic $b$ axis is higher than the corresponding energies along the crystallographic $a$ and $c$ axes (See Supplementary Fig. S3(a)). Our calculations show that the Nd sublattice exhibit biaxial magnetic anisotropy, where the easy and intermediate magnetic axes are along the crystallographic $a$ and $c$ axes, respectively and the magnetic hard axis is along the crystallographic $b$ axis (see Table ~\ref{T_SE} and Supplementary Fig. S3(a)). The calculated orbital moment of Nd $\sim$ 1.63 $\mu_B$ and is oriented antiparallel to the spin moment. The net magnetic moment $\mu_{Nd} \sim$ 1.33 $\mu_B$, lying well within the experimentally reported range of 1.30-1.93 $\mu_B$ ~\cite{magNd1,magNd2}.
We notice here $J_{Cr-Nd}^{avg}$ = 0.29 meV, comparable to
DFT estimated SIA of Nd spins, hinting into a strong interplay between the two.

Compared to the magnetic anisotropy energies of the Nd ions, the anisotropy energies of the Cr ions are order of magnitude lower. To obtain the magnetic anisotropy of the Cr ions, we switch on the LS coupling only for the Cr ions. The computed orbital moment at the Cr site $\sim$ 0.03 $\mu_B$. As shown in the Supplementary Fig. S3(b), Cr tends to orient along the crystallographic $c$ axis and this tendency enhances with the increase in $J_H$. This observation agrees with the experimentally observed stabilization of the $G_z$ magnetic phase below N\'{e}el temperature~\cite{shamir}. However, the weak magnetic anisotropy of the Cr spins is expected to be associated with high numerical error. We, therefore, scan the stability of the magnetic phases in NdCrO$_3$ as a function of the SIA parameters of the Cr spins ($E_{Cr}$ , $D_{Cr}$) employing finite temperature Monte Carlo simulations.

\subsection{Finite temperature Monte Carlo Simulations}

\begin{figure}
\includegraphics[width=1.0\linewidth]{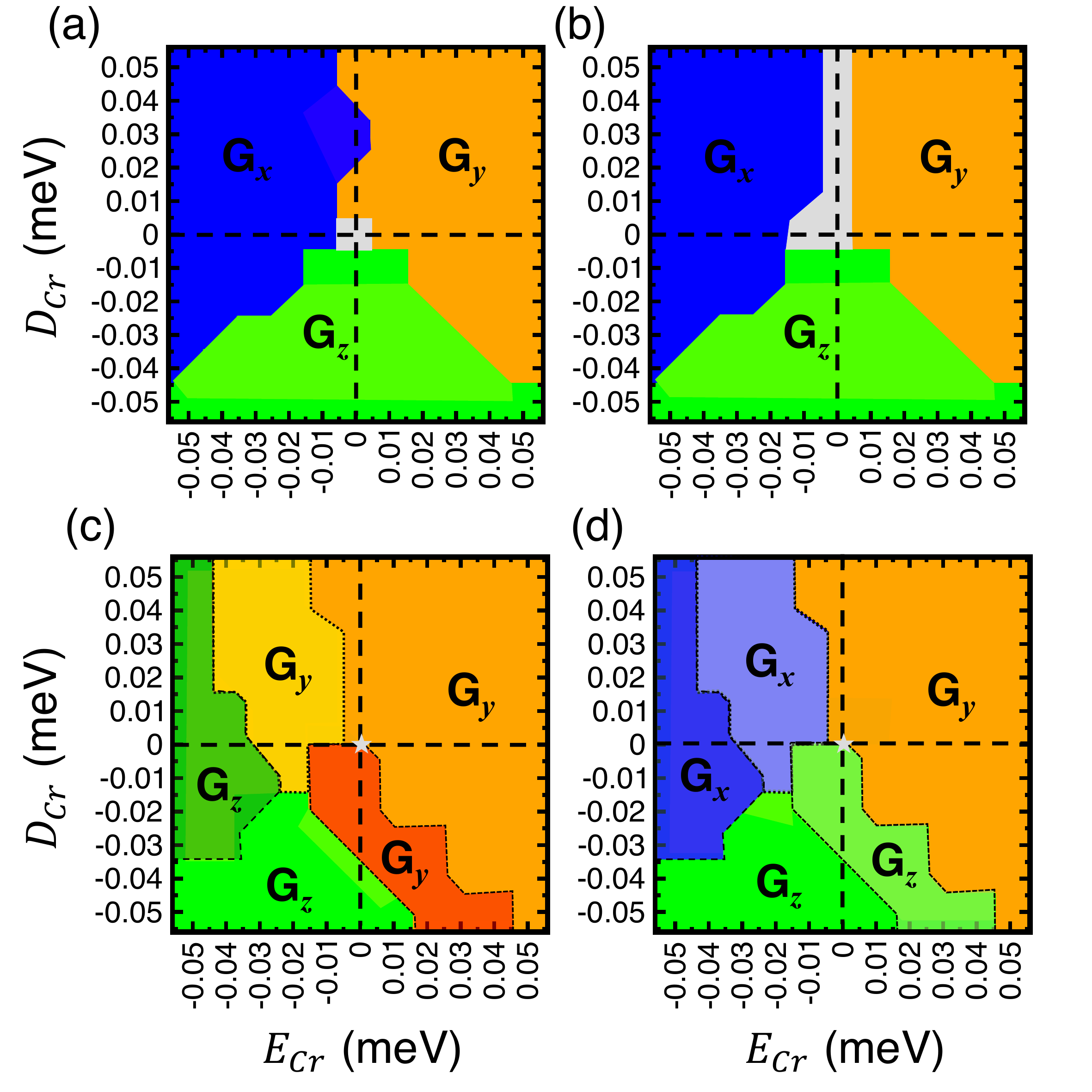}
\caption{Observed collective spin ordering in the Cr sublattice at 5 K (a) and 75 K (b) as a function of $E_{Cr}$ and $D_{Cr}$ for $|E_{Nd}|=|D_{Nd}|=0$.
  The same for GGA+$U$ estimated values of $E_{Nd}$ and $D_{Nd}$ at 5 K (c) and 75 K (d). $G_x$, $G_y$ and $G_z$ denote orientation of Cr spins along the crystallographic $a$, $b$ and $c$ axes, respectively. Different phases are shaded with different colors. The color coding is same between 5 K and 75 K, if
  there is no change in symmetry of spin ordering of Cr spins and shaded differently when there is a change in symmetry of spin ordering of Cr spins between
  5 K and 75 K. The values of isotropic magnetic exchanges are kept fixed at GGA+$U$ estimated values obtained with choice of $J_H$ = 0.3 eV. Within the parameter region shaded with light grey, none of the following phases, $G_x$, $G_y$ and $G_z$, are formed.}
\label{MC-PD}
\end{figure}

Monte Carlo (MC) simulations were performed using the model Hamiltonian constructed with the $U$ values of 2.2 eV and 5.5 eV at the Cr 3$d$ and Nd 4$f$ states, respectively and $J_H$ = 0.3 eV. This  set of particular values of $U$ and $J_H$ were taken into consideration, since the magnetic properties of the material, when calculated using this particular set, were in best agreement with the experimentally obtained results~\cite{tn1,tn2,shamir}. The primary motivation of the present study is to explore the interplay between isotropic Nd-Cr exchange interactions and SIA which can induce the spin-reorientation transitions. With this motivation in mind, we first explore the parameter space of SIA, ($E_{Nd}$, $D_{Nd}$) and  ($E_{Cr}$, $D_{Cr}$).  

\textbf{Case: $E_{Nd} = D_{Nd} =0$.}  We start our discussion by considering the MC simulation results where the magnetic anisotropy of the Nd ions is switched off ($E_{Nd} = D_{Nd} = 0$) and  the magnetic anisotropy of Cr ions is varied from -0.05 to 0.05 meV. Figure ~\ref{MC-PD}(a) and (b) show the observed magnetic order in the Cr sublattice at 5 K and 75 K, respectively, as a function of $E_{Cr}$ and $D_{Cr}$. We determined the order by calculating the staggered magnetization as defined in Eq.~\ref{AFM}. Irrespective of the value of $E_{Cr}$ and $D_{Cr}$, the specific heat, calculated as a function of temperature shows a peak at around 172 K, indicating a magnetic transition from the paramagnetic phase to a G-type AFM phase in the Cr sublattice and a disordered state in the Nd sublattice. Depending on the parameter values of
$E_{Cr}$ and $D_{Cr}$, $G_x$ or $G_y$ or $G_z$ phase is stabilized in Cr sublattice. Comparison between Figure ~\ref{MC-PD}(a) and (b) reveals
no spin-reorientation transition of the Cr spins was observed down to 5K. This provides the key finding of our study that for SR in Cr sublattice
the SIA of Nd ions plays a crucial role.\\
\begin{figure}
\includegraphics[width=1.1\linewidth]{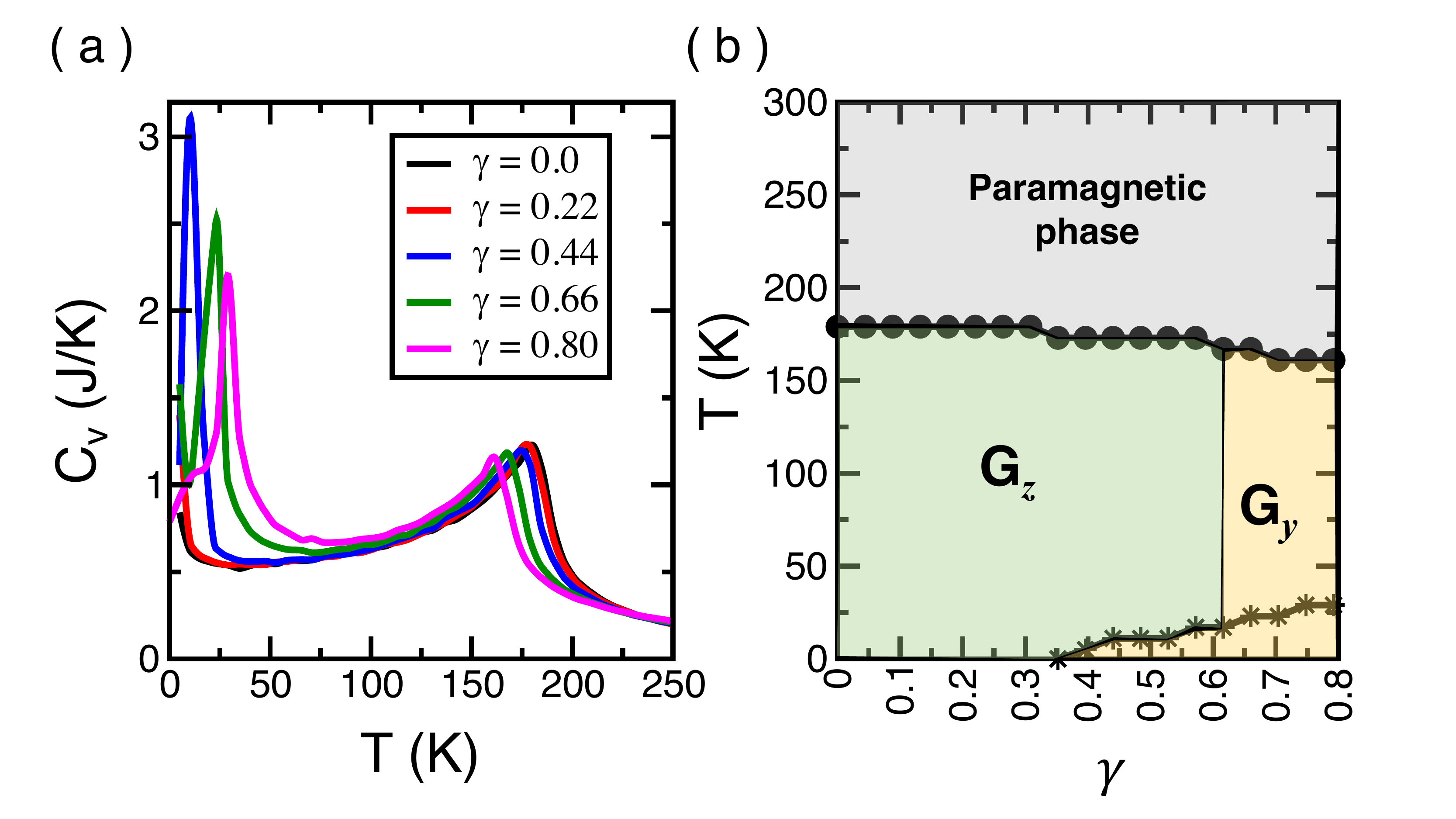}
\caption{(a) Calculated temperature dependence of specific heat for a choice of $\gamma$ values. We used $E_{Cr}$ = 0.02 meV and $D_{Cr}$ = -0.05 meV. Though the nature of magnetic phase transitions strongly depend on the magnetic anisotropy of Nd and Cr ions, the value of transition temperatures shows weak dependence. (b) Magnetic phase diagram in $T$-$\gamma$ plane,
  demarcating the paramagnetic, $G_z$ and $G_y$ spin ordered phases of Cr sublattice. The calculated N\'{e}el temperature $T_{N}$ values, plotted in solid circles, denote paramagnetic to G-type AFM ordering temperature, while the second transition temperature is plotted in stars. For $\gamma$ $\sim$ 0.35 - 0.62, stars denote
  $G_z$ $\rightarrow$ $G_y$ SR transition.}
\label{PD}
\end{figure}
\textbf{Case: $E_{Nd} \neq 0$, $D_{Nd} \neq 0$.} Figure ~\ref{MC-PD}(c) and (d) summarizes the results of MC simulations considering the DFT estimated values of SIA $E_{Nd}$ and $D_{Nd}$ parameters (see Table ~\ref{T_SE}). The main findings are as follows; (1) The calculated specific heat for a temperature
range of 300-5 K, shows two peaks. One at around 172 K and another at around 11 K. The first transition temperature corresponds to N\'{e}el temperature ($T_{N}$) forming G-type Cr spin order. $T_{N}$ value remained unchanged between $E_{Nd} = D_{Nd} =0$, and $E_{Nd} \neq 0$, $D_{Nd} \neq 0$, indicating
that high temperature magnetic ordering of Cr sublattice does not depend on Nd sublattice. However, contrary to $E_{Nd} = D_{Nd} =0$ case,
SR transition is observed in Cr sublattice depending on the choice of $E_{Cr}$ , $D_{Cr}$ (compare Figure ~\ref{MC-PD}(c) and (d)). The existence and
precise nature of the SR transition is found to strongly depend on the value of Cr SIA parameters, as depicted in various shaded regions in Figure ~\ref{MC-PD}(c) and (d). (2) For the choice of ($E_{Cr}$ , $D_{Cr}$) parameters within a critical range, as shown by the region shaded in red in Figure ~\ref{MC-PD}(c), a reorientation of the G-type ordered Cr spins from $z$ to $y$ axis ($G_z$ $\rightarrow$ $G_y$) at 11 K is observed,  which corresponds to the experimentally reported spin reorientation transition observed in NdCrO$_3$~\cite{tn2}. (3) Additionally, our results highlight a comprehensive correlation between the magnetic anisotropy of the magnetic ions and the nature of SR transitions in the Cr sublattice. The Cr spins, notably, either orient along the $y$ or the $z$ axis, below the second magnetic transition temperature. The strong tendency of the Nd spins to be oriented along the $x$ axis, exclude the formation of $G_x$ magnetic phase. 

The DFT estimated SIA coefficients of Cr ions are, $E_{Cr}$ = 0.003 meV and $D_{Cr}$ = -0.01 meV, which lie in the critical parameter space that leads to the $G_z$ $\rightarrow$ $G_y$ spin-reorientation transition at the second transition temperature. Two noteworthy points in this regard are,  (I) the $G_z$ $\rightarrow$ $G_y$ SR transition takes place for $D_{Cr}$ $<$ 0, {\it i.e.} when the magnetic easy axis of the Cr sublattice is along the crystallographic $c$ axis, which is in line with the observed trend of magnetic anisotropy of Cr ions through DFT calculations, and (II) based on our present mechanism, SR transitions occur if the magnetic easy axis of Cr sublattice coincides with one of the easy axes of Nd sublattice and due to the strong Cr-Nd superexchange interactions below SR transition temperature ($T_{SR}$) Cr spin orient along the magnetic hard axis of Nd sublattice. The calculated $T_{N}$ and $T_{SR}$ (see Figure ~\ref{PD}(a)) are in same ball park but somewhat underestimated compared to the experimentally reported values of 224 K and 34 K~\cite{tn2}, respectively. The present model, however, excludes the effect of anisotropic and anti-symmetric exchange interactions,
which can influence the precise values of the magnetic transition temperatures. The magnetic exchanges and SIA being strongly dependent on choice
of $J_H$ value provides another avenue of fine tuning of the transition temperatures, which we refrain from doing as our motivation is to unravel
the origin rather than providing a perfect matching with experimental transition temperatures.

In order to further explore the correlation between the relative strength of the exchange interactions between the two magnetic sublattices and the SR transitions, we carried out additional MC simulations as a function of $\gamma$ keeping the DFT determined relative strength of all four Cr-Nd interactions fixed. The $\gamma$ $\sim$ 0.44 case corresponds to the DFT estimated parameters. The calculated Cr sublattice staggered magnetization as a function of $\gamma$ is shown in Supplementary Fig. S4(b). Our results demonstrate existence of a second transition only above a critical value of $\gamma$ = 0.35, as shown in Figure ~\ref{PD}(a). While the value of N\'{e}el temperature decreases with the increase of $\gamma$,
the value of second transition temperature increases. As shown in the $\gamma-T$ phase diagram in
Figure ~\ref{PD}(b) and Supplementary Fig. S4(b), within a range of $\gamma$ $\sim$ 0.35 - 0.62, NdCrO$_3$ is found to to show $G_z$ $\rightarrow$ $G_y$ SR transition.
For $\gamma >$ 0.62, the second transition corresponds to the spin ordering in the Nd sublattice, and
not to SR transition of Cr sublattice.

The above exercise conclusively establishes that the SR transition of NdCrO$_3$ is a complex interplay of Nd-Cr magnetic exchanges, SIA
of Nd and Cr sublattice.

\subsection{Magnetic ordering in Nd sublattice}

So far, we have discussed the collective magnetic ordering in the Cr sublattice as observed at different temperatures. In this subsection
we take up the case of Nd sublattice.

The nature of the magnetic ordering corresponding to the Nd sublattice till date remains ambiguous~\cite{tn2,solidstate,shamir}. The
very existence of magnetic ordering in Nd  sublattice, as a matter of fact, is debated, given the fact the ordering must occur at very
low temperature due to weak Nd-Nd interaction. The evidence of ordering of R sublattice is normally
manifested as a small $\lambda$ peak in specific heat superimposed on Schottky anomaly, the latter arising due to finite R-M interaction.
In contrast to Nd ferrite, no such $\lambda$ peak is observed in the specific heat of NdCrO$_3$ suggesting cooperative ordering of Nd sublattice
is prohibited~\cite{solidstate}. Neutron diffraction study~\cite{shamir} on the other hand, suggest C-type magnetic ordering in Nd  sublattices, where
Nd spins are ordered in AFM and FM manners in the crystallographic $(001)$ plane and along the crystallographic $[001]$ axis, respectively (see Fig. 5(a)). 
We denote this magnetic structure as $C_{001}$. However, our MC results below $T_{SR}$, show the stabilization of a different variety
of C-type ordering. Here, the Nd spins form AFM ordered patterns along the crystallographic [001] and [1-10] axes and FM ordered pattern along the crystallographic [110] axis. We observe that the Nd spins are directed along [100] axis ($x$ axis) (cf Fig. 5(b)). We denote this magnetic ordering as $C_{110}$.  This
type of magnetic ordering is expected to break the crystal symmetry, leading to doubling of the size of the unit cell.

We though need to keep few points in mind. The strong isotropic exchange interactions between two magnetic sublattices create $C_{110}$ type canted
spin component in the Cr sublattice. Below $T_{SR}$ the magnetically ordered Cr sublattice is expected to create a strong anisotropic and anti-symmetric
exchange field on the Nd spin. This field can compete with the SIA tendency, and can influence the precise nature of spin ordering of Nd sublattice. 
Our present calculation does not take into account these additional factors which may influence the magnetic order in the Nd sublattice at
low temperatures, leaving the issue open. Further investigations need to be conducted to gain a deeper insight in this regard.  
\begin{figure}
\includegraphics[scale=0.25]{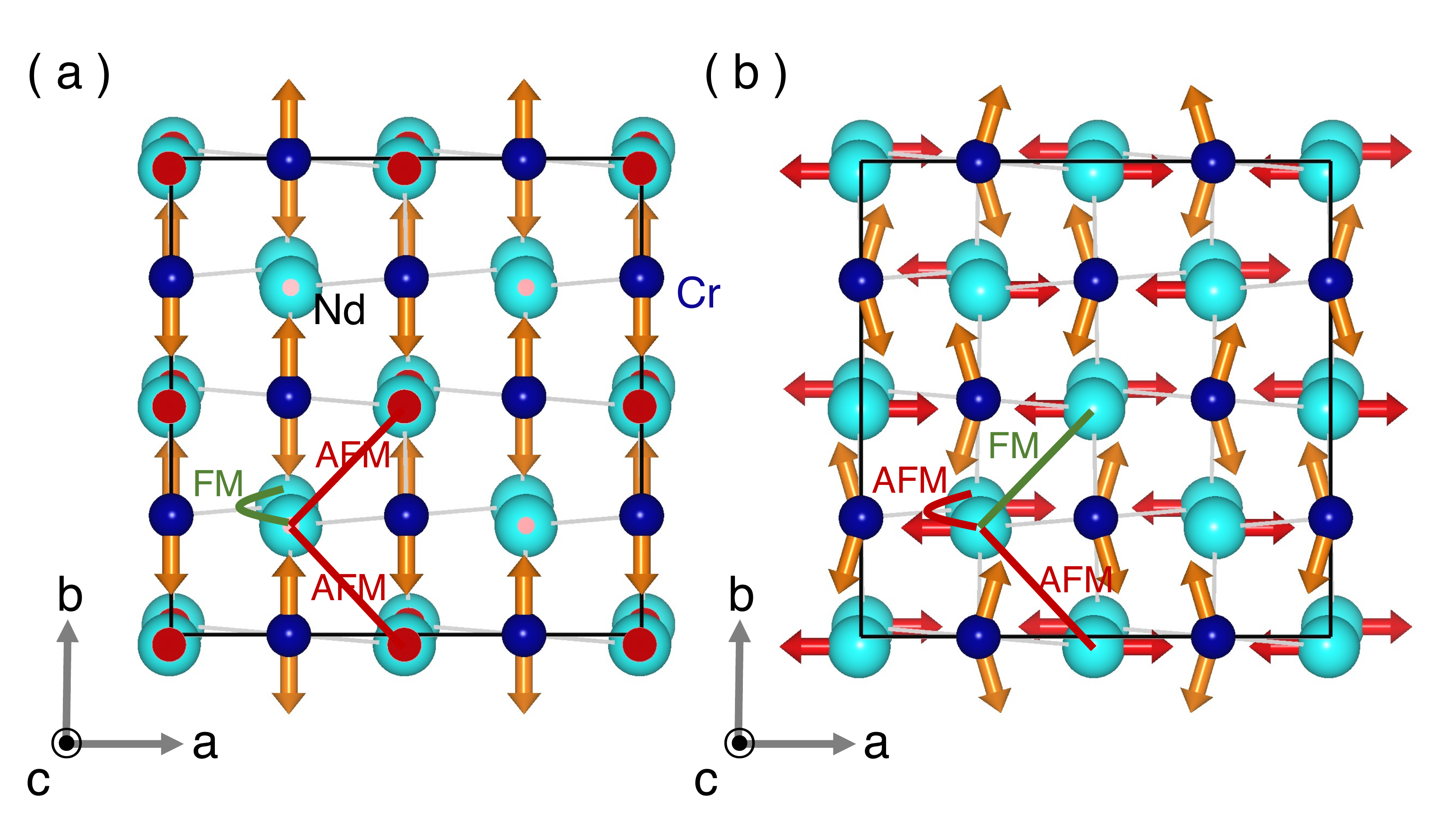} 
\caption{(a) Experimentally proposed $C$-type magnetic ordering in the Nd sublattice \cite{shamir} having $Pbnm$ magnetic symmetry. Nd spins (denoted by red arrow) orient along $z$ axis. The Nd spins order in AFM and FM patterns in the $(001)$ plane and along the $[001]$ axis, respectively. We denote this magnetic ordering as $C_{001}$. (b) Proposed $C$-type magnetic ordering in the Nd sublattice through the MC simulations. Nd spins are oriented along $x$ axis. The Nd spins form AFM ordering along the [001] and [1-10] axes and FM ordering along [110] axis. We denote this magnetic ordering as $C_{110}$. The Cr spins (denoted by orange arrow) are ordered in $G_y$ pattern. The strong isotropic exchange interactions between two magnetic sublattices create $C_{110}$ type canted spin component in the Cr sublattice.}
\label{MAG}
\end{figure}

\section{CONCLUSION \& DISCUSSIONS}

Using combination of first-principles calculations and finite-temperature MC simulations on DFT derived spin Hamiltonian we studied
the complex magnetism of NdCrO$_3$, especially the spin reorientation transition of Cr sublattice. The most exhaustive theoretical study
on spin reorientation in rare-earth orthochromites and orthoferrites~\cite{yamaguchi}, given about three decades ago,  was based on mean field study of the spin Hamiltonian
in parameter space. While this study was successful in explaining SR of both category (I) and (II), it failed to explain SR of category
(V) {\it i.e} that of NdCrO$_3$ motivating the present study. Our study importantly took into account the SIA of Nd spins, which was neglected
in the study by Yamaguchi~\cite{yamaguchi}. In order to gain understanding on the microscopic origin of SR transition in NdCrO$_3$, we carried
out calculations switching off and on SIA of Nd spins, varying the SIA of Cr spins around the DFT estimated values, and varying the Nd-Cr
exchange strength, parameterized through $\gamma$.

Our analysis established the crucial role of SIA of Nd spins in driving the SR transition, provided
the Nd-Cr magnetic exchange is above a critical strength. Strengthening of Nd-Cr exchange further beyond another limit leads to
cooperative ordering of Nd sublattice avoiding the SR transition of Cr sublattice. Furthermore, the experimentally observed nature of SR
transition in NdCrO$_3$ was found to be reproduced only for a restricted parameter range of SIA of Cr spins, highlighting of
importance of SIA of Cr spins, though it is order of magnitude weaker in strength compared to SIA of
Nd spins. Thus, our findings point to the fact that SR observed in NdCrO$_3$ arises out of delicate balance between Nd-Cr exchange,
Nd SIA as well as Cr SIA. While the significant strength of Nd-Cr exchange and SIA of Nd spins set the stage,
the SIA of Cr spins which is an order of magnitude smaller compared to that of Nd spins, decides the details.

We would like to end our discussion with some open issues and suggestions. Our study as mentioned already did not take into
account the anti-symmetric and anisotropic-symmetric exchange interactions, rather focused only on isotropic exchanges.
The anti-symmetric and anisotropic-symmetric exchanges in M sublattice are expected to give rise to canting and ferromagnetic component to Cr spin ordering, as
observed experimentally. These exchanges between M and R sublattices were suggested to be crucial for driving SR in GdCrO$_3$
for example, though they are expected to be one and two orders of magnitude smaller compared to isotropic exchanges. We
note the importance of SIA is negligible for Gd$^{3+}$ thus making the anti-symmetric exchange of M-R interaction
important. In case of NdCrO$_3$, however, consideration of only the isotropic M-R exchange in presence of finite, and moderately
strong SIA of Nd spins, together with weak but finite SIA of Cr spins, was sufficient to capture the SR. The effect of
anti-symmetric and anisotropic-symmetric exchange interactions, thus would come as secondary effect. 

While our study primarily focused on SR transition, we also studied the possibility of cooperative ordering of Nd spins, which remains
controversial. Our study points to a $C_{110}$ type ordering of Nd spins, in contrast to $C_{001}$ type ordering suggested from neutron
diffraction~\cite{shamir}. This will be taken up in a future study involving a more complete Hamiltonian. Also it calls for further
experimental studies.

Finally, it will be an interesting idea to study the SR in mixed orthoferrite-chromite compound, NdFe$_{x}$Cr$_{1-x}$O$_3$, given
the fact that $\gamma$ is about a factor of two smaller in NdFeO$_3$ compared to NdCrO$_3$, while the SIA of Fe is expected to
be smaller compared to Cr due to its suppressed orbital degeneracy. This may make the influence of
the anti-symmetric and anisotropic-symmetric exchange interactions important.  It is interesting to note that both the high temperature
and the low temperature symmetry of the spin ordering of M sublattice in NdFeO$_3$ is different from that of NdCrO$_3$, it being
$G_x$ ($G_z$) for NdFeO$_3$ and $G_z$($G_y$) for NdCrO$_3$ before (after) SR. To the best of our knowledge, the phase
diagram of mixed compound, NdFe$_{x}$Cr$_{1-x}$O$_3$ is yet to be explored either experimentally or theoretically. We hope our study
will motivate further studies in this direction.

\section{ACKNOWLEDGEMENTS}

The authors gratefully acknowledge discussion with Badiur Rahaman, Sudipta Bandyopadhyay and Sourav Kanthal. Research at Tokyo Institute of Technology is supported by the Grant-in- Aid for Scientific Research 19K05246 and 19H05625 from the Japan Society for the Promotion of Science (JSPS). HD acknowledges computational support from TSUBAME supercomputing facility. TS-D acknowledges J.C.Bose National Fellowship (grant no.JCB/2020/000004) for funding. 

%\subsection{Single-ion Anisotropies}
%
%\section{MC Study of Spin-reorientation}
%
%\subsection{Influence of Single-ion Anisotropy}
%
%\begin{figure}
%\includegraphics[width=1.0\linewidth]{Fig4.jpg}
%\caption{}
%\end{figure}
%
%\subsection{Influence of Nd-Cr interaction}
%
%\begin{figure}
%\includegraphics[width=1.0\linewidth]{Fig5.jpg}
%\caption{}
%\end{figure}
%
%\begin{figure}
%\includegraphics[width=1.0\linewidth]{Fig6.jpg}
%\caption{}
%\end{figure}

%\section{Supplementary Materials} 
\newcommand{\beginsupplement}{%
        \setcounter{table}{0}
        \renewcommand{\thetable}{S\arabic{table}}%
        \setcounter{figure}{0}
        \renewcommand{\thefigure}{S\arabic{figure}}%
     }

\beginsupplement

\begin{figure*}
\includegraphics[width=0.8\linewidth]{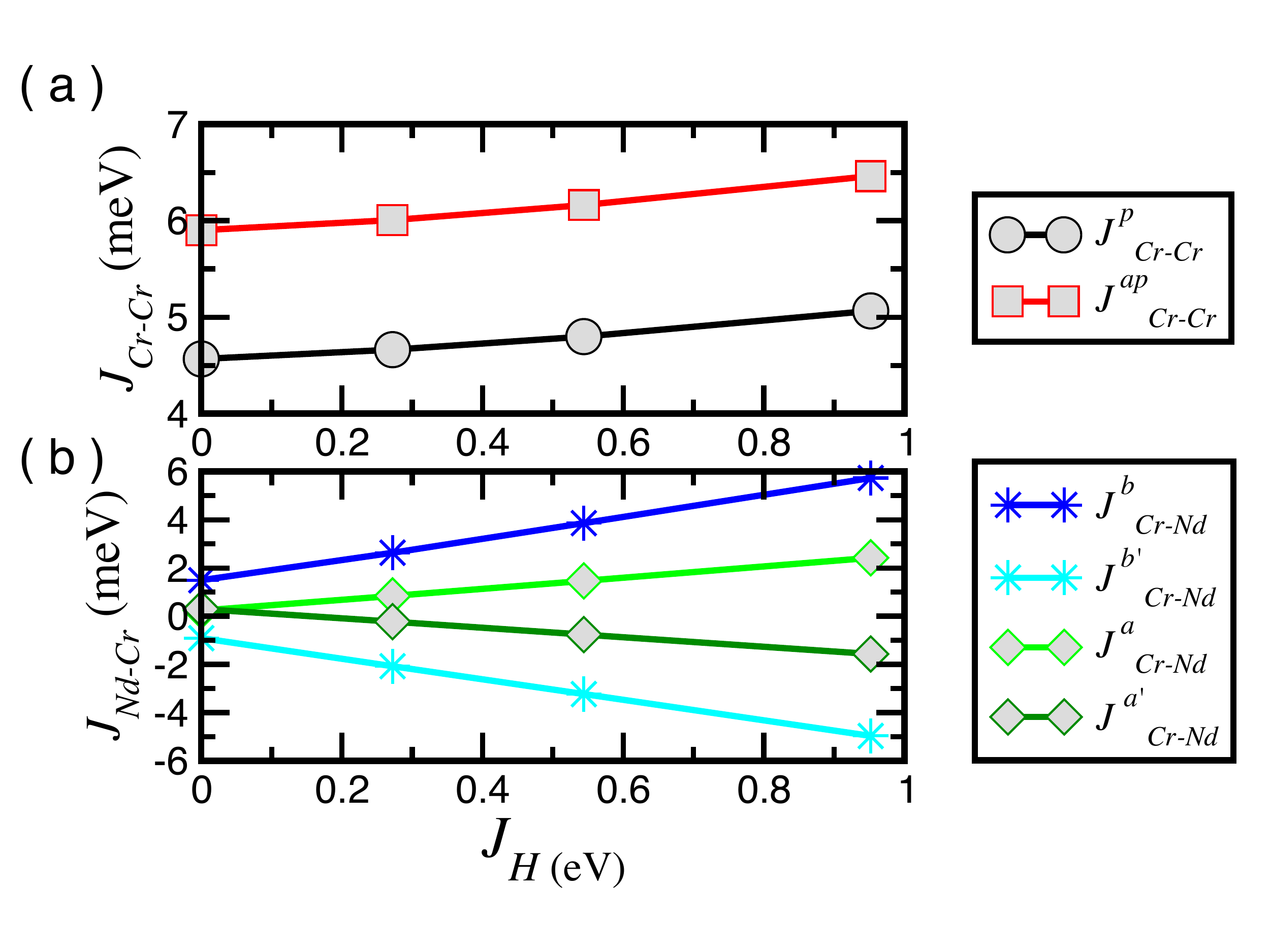}
\caption{Calculated superexchange interactions between Cr spins (a) and Cr and Nd spins (b) as a function of $J_H$. We present the results for the $U$ value of 2.2 eV and 5.5 eV at Cr 3$d$ and Nd 4$f$ site, respectively.}
\label{NCO-J}
\end{figure*}

\begin{figure*}
\includegraphics[width=0.8\linewidth]{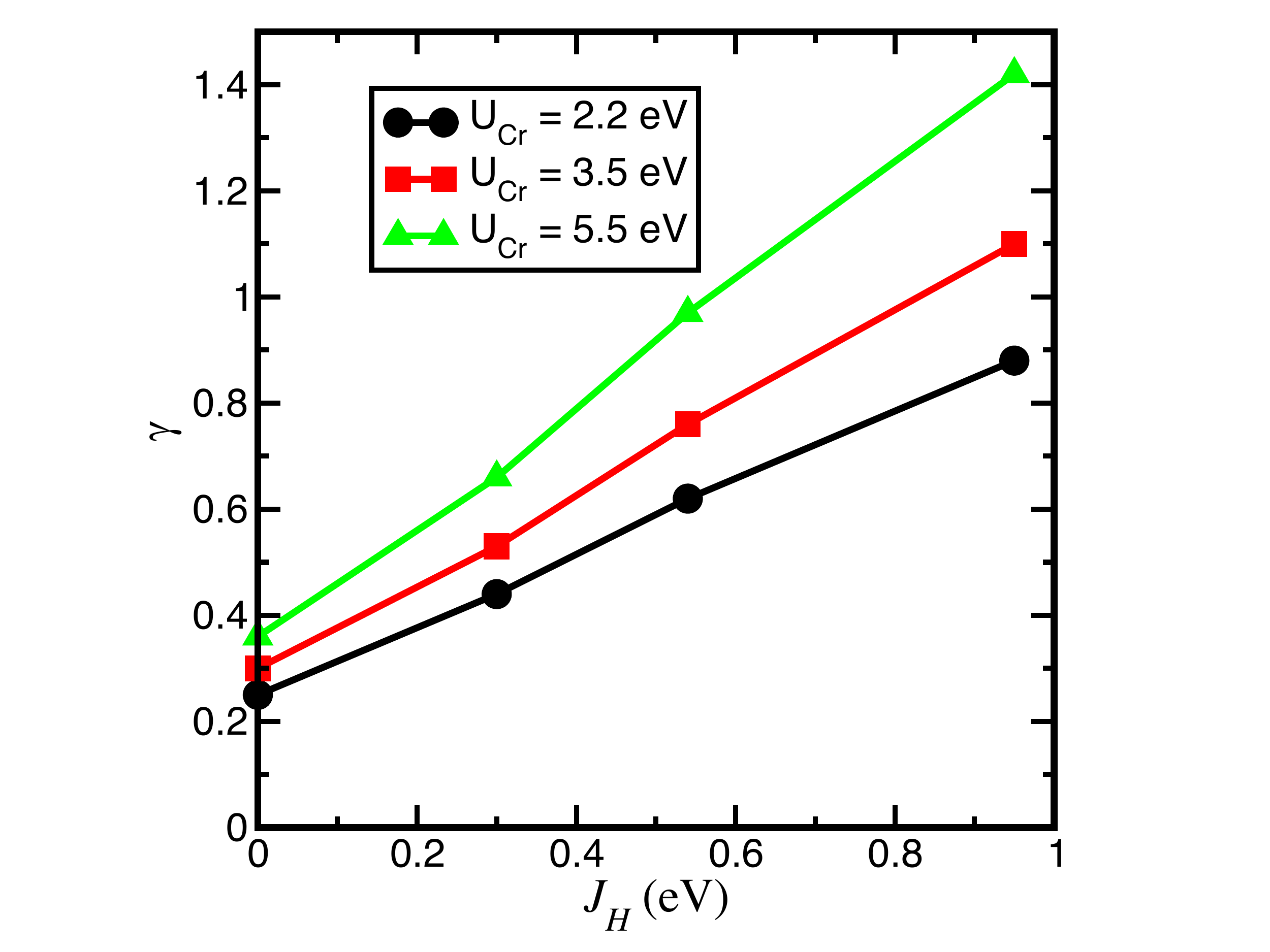}
\caption{Estimated relative strength of Cr-Nd magnetic interaction with respect to Cr-Cr magnetic interaction, which is denoted as $\gamma$, as a function of $J_H$ and $U$ value of Cr. We used $U$ = 5.5 eV value for Nd 4$f$ site.}
\label{NCO-Gamma}
\end{figure*}
 
\begin{figure*}
\includegraphics[width=1.0\linewidth]{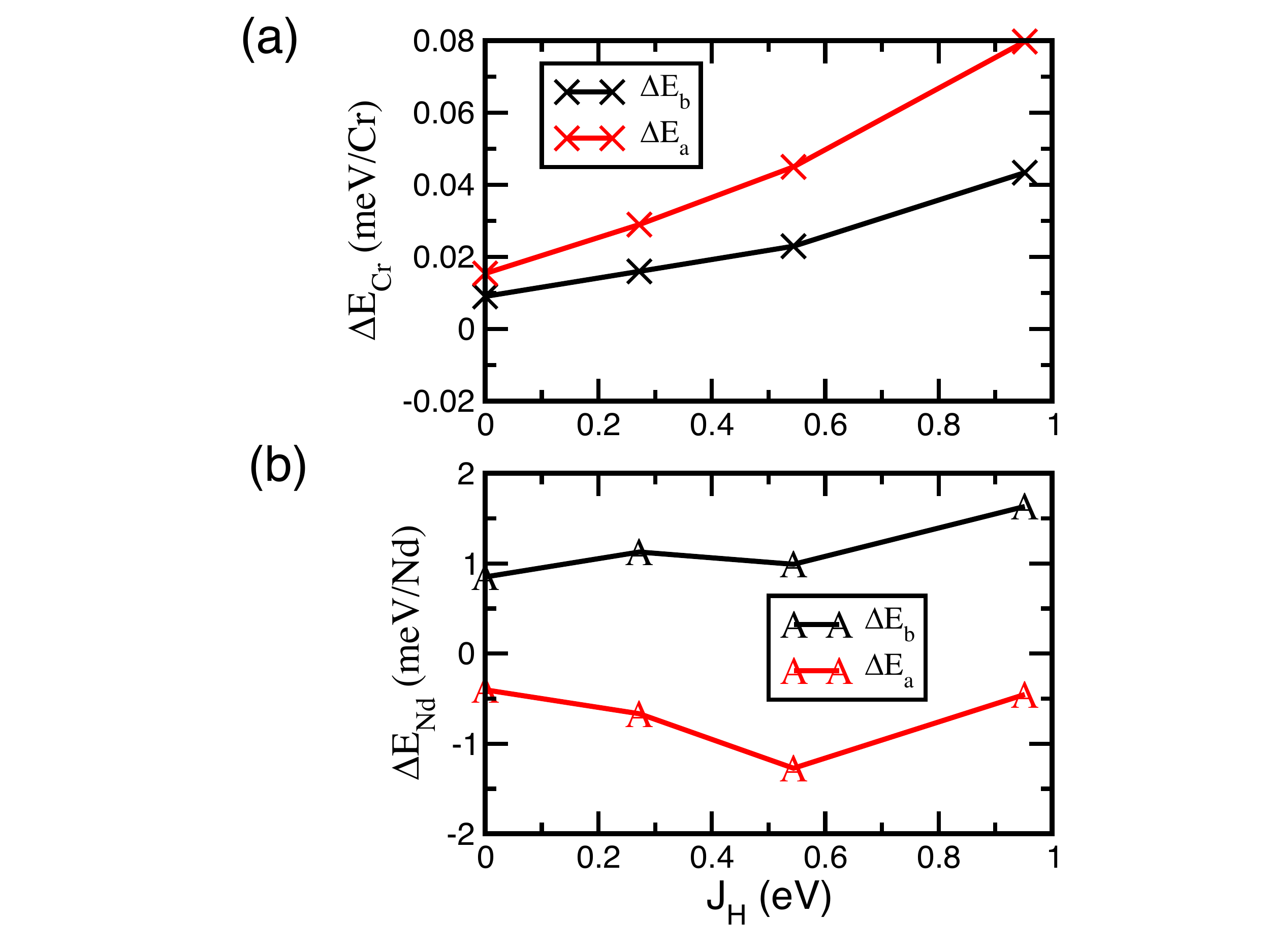}
\caption{Calculated relative magnetic anisotropy energies (MAE) for the Cr (a) and Nd (b) ions, as a function of $J_H$. The relative MAE are defined as, $\Delta E_{b}=E_{b}-E_{c}$ and $\Delta E_{a}=E_{a}-E_{c}$. Where $E_{a}$, $E_{b}$ and $E_{c}$ denote total energy corresponding to the spin orientation along crystallographic $a$, $b$ and $c$ axes, respectively. We present the results for the $U$ value of 2.2 eV and 5.5 eV at Cr 3$d$ and Nd 4$f$ site, respectively.}
\label{NCO-SIA}
\end{figure*}

\begin{figure*}
\includegraphics[width=1.0\linewidth]{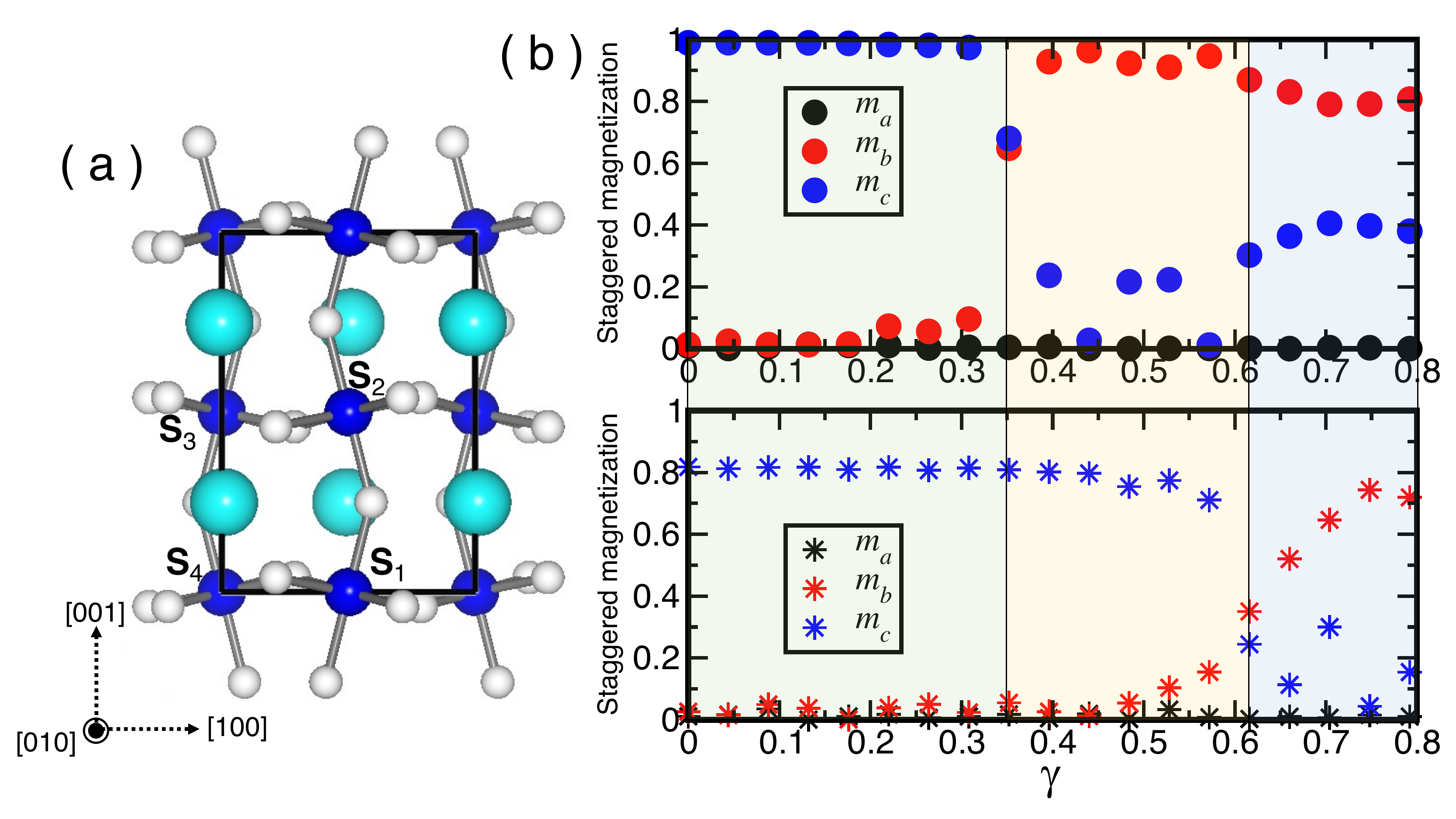}
\caption{(a) $Pbnm$ crystal structure and the Cr spins are denoted as {\bf S}$_1$$\rightarrow${\bf S}$_4$. (b) Calculated staggered magnetization at 5 K (upper panel) and 75 K (lower panel) as a function of $\gamma$.}
\label{NCO-SIA}
\end{figure*}
 

\begin{thebibliography}{99}

\bibitem{RFO-MAG}D. Treves, Phys. Rev. 125, 1843 (1962).
\bibitem{ref2} K. W. Blazey and G. Burns,  Proc. Phys. Soc. \textbf{91}, 640 (1967).
\bibitem{ref1} K. Tsushima, K. Aoyagi and S. Sugano, J. appl. Phys. {\bf 41}, 1238 (1970).

\bibitem{ref5}  R. L. White, J. Appl. Phys. 40. 1061 (1969).

\bibitem{tokura}Y. Tokunaga, S. Iguchi, T. Arima, and Y. Tokura, Phys. Rev. Lett. {\bf 101}, 097205 (2008).
\bibitem{Rao}B. Rajeswaran, D. I. Khomskii, A. K. Zvezdin, C. N. R. Rao, and A. Sundaresan, Phys. Rev. B {\bf 86}, 214409 (2012).
\bibitem{Tokunaga}Y. Tokunaga, N. Furukawa, H. Sakai, Y. Taguchi, T. Arima and Y. Tokura, Nat. Mater. {\bf 8}, 558–562 (2009).
\bibitem{Zhao}H. J. Zhao, L. Bellaiche, X. M. Chen and J. \'{I}\~{n}iguez, Nat. Commun. 8, 14025 (2017).
\bibitem{DAS}X. Ye, J. Zhao, H. Das, D. Sheptyakov, J. Yang, Y. Sakai, H. Hojo, Z. Liu, L. Zhou, L. Cao, T. Nishikubo, S. Wakazaki, C. Dong, X. Wang, Z. Hu, H.-J. Lin, C.-T. Chen, C. Sahle, A. Efiminko, H. Cao, S. Calder, K. Mibu, M. Kenzelmann, L. H. Tjeng, R. Yu, M. Azuma, C. Jin and Y. Long, Nat. Commun. {\bf 12}, 1917 (2021).
\bibitem{RFO1}M. Marezio, J. P. Remeika and P. D. Dernier, Acta Cryst. \textbf{B26}, 2008-2022 (1970).
\bibitem{RCO1}M. C. Weber, J. Kreisel, P. A. Thomas, M. Newton, K. Sardar, and R. I. Walton, Phys. Rev. B {\bf 85}, 054303 (2012).
\bibitem{yamaguchi} T. Yamaguchi, J. Phys. Chem. Solids. {\bf 35}, 479 (1974).
\bibitem{MF-RMO}E. Bousquet and A. Cano, J. Phys.: Condens. Matter \textbf{28}, 123001 (2016).
\bibitem{NFO-TN}D. Treves, J. Appl. Phys. \textbf{36}, 1033 (1965).
\bibitem{tn1}  E. F. Bertaut, J. Mareschal, G. de Vries, R. Aleonard, R. Pau-
  thenet, J. P. Rebouillat, and J. Sivardiere, IEEE Trans. Magn. {\bf 2}, 453 (1966).
\bibitem{tn2} F. Bartolom\'{e}, J. Bartolom\'{e},  M. Castro, and J. J. Melero, Phys. Rev. B {\bf 62}, 1058 (2000).
\bibitem{DM1}I. Dzyaloshinski, J. Phys. Chem. Solids 4, 241 (1958). 
\bibitem{DM2}  T. Moriya, Phys. Rev. 120, 91 (1960); In: {\it Magnetism I} (Edited by G. T. Rado and H. Suhl), p. 85. Academic Press, New York
  ( 1963).

\bibitem{shamir} N. Shamir, H. Shaked, S. Shtrlkman, Phys. Rev. B {\bf 24}, 6642 (1981).
\bibitem{bertaut} E. F. Bertaut and J. Mareschal, Solid. Stat. Commun., {\bf 5}, 93, (1967).  
\bibitem{horneich-old} R. M. Hornreich, Y. Komet, R. Nolan, B. M. Wanklyn, I. Yaeger, Phys. Rev. B {\bf 12}, 5094 (1975).
\bibitem{solidstate} F. Bartolom\'{e}, M. D. Kuzmin, J. Bartolom\'{e}, J. Blasco, J. Garc\'{i}a, and F. Sapi\~{e}a, Solid State Commun. 91, 177 (1994).
\bibitem{jalcom} J.Ramesh, N.Raju, S.Shravan Kumar Reddy, M.Sreenath Reddy, Ch.Gopal Reddy, P.Yadagiri Reddy, K.Rama Reddy,
  V.Raghavendra Reddy, J. Alloy Compounds, {\bf 711} 300 (2017).
\bibitem{horneich}  R.M. Hornreich, J. Magn. Magn Mater. {\bf 7} 280 (1978).
\bibitem{ref11-prb} J. B. Ayasse, A. Berton, and J. Sivardiere, C. R. Seances Acad. Sci., Ser. B 271, 1220 (1970).

\bibitem{wien2k1} P. Blaha, K. Schwarz, G. K. H. Madsen, D. Kvasnicka, J. Luitz, R. Laskowski, F. Tran and L. D. Marks, WIEN2k: An Augmented Plane Wave plus Local Orbitals Program for Calculating Crystal Properties; Vienna University of Technology: Austria, 2018 \url{http://www.wien2k.at/index.html}.
\bibitem{wien2k2}P. Blaha, K.Schwarz, F. Tran, R. Laskowski, G.K.H. Madsen and L.D. Marks, J. Chem. Phys. \textbf{152}, 074101 (2020).
\bibitem{PBE}J. P. Perdew, K. Burke, and M. Ernzerhof, Phys. Rev. Lett. \textbf{77}, 3865 (1996).
\bibitem{DFT+U}V. I. Anisimov, I. V. Solovyev, M. A. Korotin, M. T. Czy\.{z}yk, and G. A. Sawatzky, Phys. Rev. B, {\bf 48}, 16929 (1993).
\bibitem{cDFT} P. Dederichs, S. Bl\"{u}gel, R. Zeller, H. Akai, Phys. Rev. Lett. {\bf 53}, 2512 (1984).
\bibitem{U-cr} V. I. Anisimov, J. Zaanen, and O. K. Andersen, Phys. Rev. B {\bf 44}, 943 (1991); W. E. Pickett, S. C. Erwin, and E. C. Ethridge
Phys. Rev. B {\bf 58}, 1201 (1998).
\bibitem{structure}J. Prado-Gonjal, R. Schmidt, J.-J. Romero, D. \'{A}vila, U. Amador, and E. Mor\'{a}n, Inorg. Chem. {\bf 52}, 1, 313–320 (2013).
\bibitem{magNd1}E. F. Bertaut and J. Mareschal, Solid State Commun. {\bf 5}, 93 (1967).
\bibitem{magNd2}N. Shamir, H. Shaked, and S. Shtrikman, Phys. Rev. B {\bf 24}, 6642 (1981).

\end{thebibliography}
\end{document}